\documentstyle[12pt,epsf]{article} 

\newcommand{\bi}{\begin{itemize}}
\newcommand{\ei}{\end{itemize}}
\newcommand{\bc}{\begin{center}}
\newcommand{\ec}{\end{center}}

\newcommand{\be}{\begin{equation}}
\newcommand{\ee}{\end{equation}}

\newcommand{\bqn}{\begin{eqnarray}}
\newcommand{\eqn}{\end{eqnarray}}

\begin{document}

\title{ CONSTRUCTING BIDIMENSIONAL SCALAR FIELD THEORY MODELS FROM ZERO MODE
FLUCTUATIONS} 

\author{Gabriel H. Flores$\;$\thanks{E-mail:gflores@cbpf.br} and
N. F. Svaiter$\;$\thanks{E-mail: nfuxsvai@cbpf.br}\\ \\
{\it  Centro Brasileiro de Pesquisas Fisicas-CBPF,}\\
{\it Rua Dr.Xavier Sigaud 150, 22290-180 Rio de Janeiro, RJ, Brazil} }

\maketitle

\begin{abstract} 
In this paper we review how to reconstruct scalar field theories in
two dimensional spacetime starting from solvable Schrodinger equations.
Three different Schrodinger potentials are analysed. We obtained two 
new models starting from the Morse and Scarf II hyperbolic potentials,
{\it i.e}, the $U(\phi)=\phi^2\ln^2(\phi^2)$ model and $U(\phi)=\phi^2\cos^2(\ln(\phi^2))$ model
respectively.   
\vspace{0.34cm} 
\noindent \\
PACS number(s):11.30.Pb, 03.65.Fd, 11.10.Ef.    
\end{abstract} 

\newpage
\section{Introduction}
~~Solitons are  solutions of nonlinear equations that have the following fundamental 
properties: the profile is stable, the energy associated with them
is finite and also they behave as particles in the sense that multi-solitonic
solutions behave as independent one-soliton solutions as time goes to infinity
\cite{rajaraman}.  Also, there
are a less restricted class of solutions for non-linear equations
that have the same properties of solitonic solutions, except the property 
that they retain their shape after collision. In this case such solutions are called
solitary waves. In general solitons can exist in any (d+1) dimensional 
spacetime.
In the (1+1) dimensional case, the static solutions are called kinks. These
solutions link two degenerate trivial vacua of the theory. An important
property are that these solutions are still stable, if we take into account 
quantum corrections. A deep analysis of the quantum properties of these
solutions were carried out in the seventies. See for example ref.\cite{jackiw}
and references therein. On the other hand, there are solutions that become
unstable when quantum corrections are taking into account. In the
(1+1) dimensional case these static solutions are called lumps or bounces.

In this paper we reconstruct kink and lump profiles and the scalar field theoretical 
models that support such kink or lump like solutions starting from exactly
solvable Schrodinger equations. We use the fact that quantum
corrections around these classical solutions are given by one dimensional
Schrodinger equations. The zero mode eigenfuntion of this equations is
proportional to the derivative of the kink (or lump). This fact is based
on traslational invariance of the field theory model. Also Bogomol'nyi s         condition  give us a relation between the density potential and the zero mode eigenfuntion.
Then solving for the kink (or lump) from the zero mode and replacing in the
Bogomol'nyi condition we recover the density potential as a function of the kink
(or lump). This is a general strategy to recover the field theoretic model from
arbitrary Schrodinger potentials knowing the zero eigenfuntions. Since we
will be interested in computing quantum corrections of the field
theoretic models, we must study only exactly solvable Schrodinger equations
since in this case we are able to perform calculations of the quantum
corrections. For our knowledgement, the first authors that stressed this fact
were Christ and Lee \cite{lee}. More recently using supersymetric quantum mechanic
Casahorran {\it et al} \cite{cas1} and also Boya and Casahorran \cite{cas2}
continue this research program. For other interesting references see
for instance \cite{cas1},\cite{cas2} and \cite{lima}. We would like to stress that previously to these works, Kumar \cite{kumar} sugested
the construction of solitonic profiles using isospectral Hamiltonians. 

In (1+1) dimensions there are an infinite number of renormalizable scalar
field theory models.  Nevertheless in the literature it was considered only 
a very restricted number of them. This fact can be understood by the
following reason: only in a few of them we can go beyond the perturbative analysis. 
There is a spread opinion that in two-dimensional models one can test ideas that can 
be generalized afterward in the more interesting  (3+1) dimensional case. 
In (1+1)
dimensions it was obtained amazing results as for example the 
fractionization of charge \cite{rebi} or the emergence of fermions from
bosons \cite{coleman1}, both phenomena that have counterpart in the 
(3+1) dimensional case. 

From the classical solutions of the non-linear field equations,
we can go further and obtain quantum corrections. We have to expand
the field operator around the classical solution, and retain only quadratic 
terms (where the higher order terms can be treated perturbatively). Then, we    will
obtain a Schrodinger equation that describes the mode oscillations around
the classical solution. This Schrodinger equation admit a zero mode solution 
with eigenfunction equal to the derivative of the classical solution. 
In the case in which the classical solution is a kink, we have that this zero mode 
have the lowest eigenvalue and then all the mode
oscillations are real, that is, we have stability. In the case in which the 
classical solution is a bounce, there will be another mode solution with
negative eigenvalue, and consequently there will be an imaginary mode, that signalize
instability. As the next step, we can solve the Schrodinger equation  for all
mode solutions and  compute the first quantum corrections. In the case of kinks,
one of the quantities of physical interest is the quantum corrections to the mass of such kinks,
that is, we compute the "zero point" energy of this configuration. In
the case of bounces we compute the decay rate of the unstable vaccum.
Any way, in both cases we have to solve a Schrodinger equation that
in general can not be solved analyticaly. In the case of sine-Gordon and $\phi^4$ stable models, the Schrodinger 
equations are respectively the $N=1,~N=2$ cases of the general solvable
reflectionless potential $V(x)=-N(N+1)/\cosh^2(x)$. For the most simple
polinomial unstable $\phi^3$ model we have the $N=3$ case. This last model
was used as a laboratory for computing the decay rate (or the life time)
of a system trapped in a false vacuum \cite{IJMPA-5-1319}. Also, recently
the $\phi^3$ model was used as an exactly solvable toy model for tachyon
condensation in string field theory \cite{zwiebach}. 
For the case of the $\phi^6$ model, we have to deal with a complicated
Schrodinger equation, that can be reduced to a Heun equation \cite{bate}.
Unfortunately this equation can not be exactly integrated. Then, we see that althought 
in two dimensions
we can have an infinite number of renormalizable scalar field theories,
only for a very restricted number of them we can perform an analytic treatment in the non-perturbative sector. 

The organization of the paper is the following. In section II we show how to reconstruct
salar field theory models from zero mode solutions. In section III we analyze 
the models that arise from the Rosen-Morse II hyperbolic potential. In section
IV and V we perform the same analysis for the the Morse and the Scarf II hyperbolic potentials respectively. Conclusions are given in section VI. We use throughout this paper $\hbar=c=1$.

\section{Reconstructing the field theory models}
\begin{figure}
\centerline {\epsfxsize=3.5in\epsfysize=1.5in\epsffile{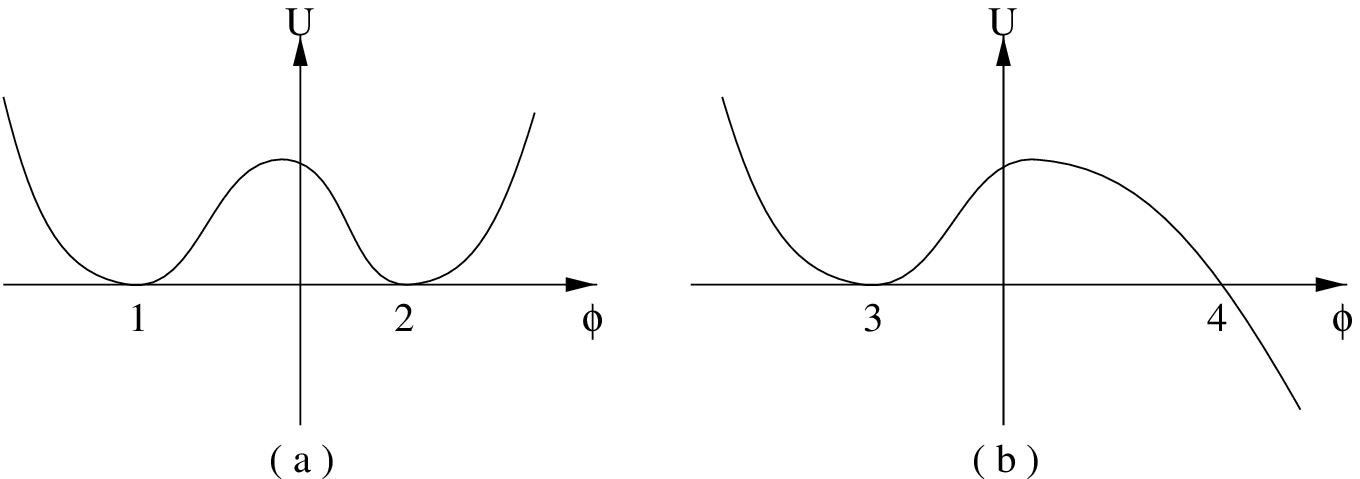} }
\caption{(a) $U(\phi)$ with two degenerate vacua and (b) with a false vacuum.}
\begin{picture}(10,10)
\end{picture}
\label{potential}
\end{figure}

In this section we briefly review how to construct scalar field theory models starting
from solvable Schrodinger equations. We start from a lagrangean 
\be
L=\int dx \left( \frac{1}{2}\partial_\mu\phi~\partial^\mu\phi-U(\phi)\right)
\;,
\label{lagran}
\ee
where $\mu=0,1$ and $U(\phi)$ with at least two degenerate absolute minima
as showed in fig.(\ref{potential}-a) or with a local minima (a false vacuum)
as showed in fig.(\ref{potential}-b). The classical equation of motion for static 
configurations are given from eq.(\ref{lagran})
\be
\frac{d^2\phi}{dx^2}=U'(\phi)\;.
\label{classic}
\ee
The eq.(\ref{classic}) can be analyzed making use of a particle mechanical analogy. 
Suppose that  $\phi$ describe the position of a particle and $x$ is the
time. Consequentely, eq.(\ref{classic}) is the equation of motion of 
a particle in a
conservative potential $-U(\phi)$. In order to analyze eq.(\ref{classic}) we have to take
into account only the possible trajectories of the 'particle' in the inverted potential. 
Clearly, we are interested only in solutions with a finite energy, in other words, 
solutions that have a finite interval of motion in $\phi$ but that are not 
oscillatory.
From the inverted potential $-U(\phi)$ given by fig.(\ref{potential}-a) it is easy to see 
that such requirement is satisfied only for those motion that take place between
the absolute minima given by points $1$ and $2$. Using the same argument for the case given by 
fig.(\ref{potential}-b) we see that we allow the motion that
start in point $3$ bounce in $4$ and return to the point $3$. In the first
case the static solution is know as  kink, while in the second case
such solution is called a lump or a bounce. From figures (\ref{potential}-a) and 
(\ref{potential}-b) we see that these solutions are integrals of motion with zero
energy, using the particle mechanical
analogy. From eq.(\ref{classic}) we have
\be
\frac{1}{2}\left(\frac{d\phi}{dx}\right)^2=U(\phi)\;,
\label{bogo}
\ee
an equation that is known as Bogomol'nyi condition \cite{bogo}. From eq.(\ref{bogo}) it
is straightforward  to obtain the kinks or lumps (we denote them as $\phi_c$) solving 
the integral
\be
x-x_0=\int^{\phi_c(x)}\frac{1}{\sqrt{2U(\phi)}}d\phi
\ee
and inverting it. 
With the classical static configuration  we can go to the first quantum 
corrections. For such purpose, we expand the time dependent field $\phi(x,t)$
around the static configuration, {\it i.e} $\phi(x,t)=\phi_c(x)+\eta(x,t)$.
Substituting this expansion in eq.(\ref{lagran}) and retaining only quadratic terms
in $\eta$ we obtain the following lagrangean:
\be
L=L[\phi_c]+\int dx\left[ \frac{1}{2}\frac{d^2\eta}{dt^2}-\frac{1}{2}\eta
\left(-\frac{d^2}{dx^2}+U''(\phi_c(x))\right)\eta\right]\;.
\label{expand}
\ee
As a next step, we use the expansion  $\eta(x,t)=\sum_n q_n(t)\psi_n(x)$,
and choosing a complete basis $\{\psi_n\}$ as solutions of the Schrodinger
equation
\be
\left( -\frac{d^2}{dx^2}+U''(\phi_c(x))\right)\psi_n(x)=\omega_n^2\psi_n(x)\;,
\label{estab}
\ee
we reduce the lagrangean given by eq.(\ref{expand}) to 
\be
L=L[\phi_c]+\sum_n(\frac{1}{2}q_n^2-\frac{\omega_n^2}{2}q_n^2)\;.
\label{sum}
\ee
From eq.(\ref{sum}) we see that the problem was reduced to a system with infinitely
uncoupled harmonic  oscillators. Now, the quantization program can be implemented in
the standard way. In particular, the zero point energy,  that in the
case of kinks are interpreted as its mass \cite{rajaraman}, is given by
\be
H=H[\phi_c]+\frac{1}{2}\sum_n\omega_n\;.
\label{massa}
\ee
Taking the derivative of eq.(\ref{bogo}) it is easy to see that eq.(\ref{estab}) 
always admit a solution with eigenvalue $\omega_0^2=0$ and with respective
eigenfuntion $\psi_0$ given by 
\be
\psi_0\propto\frac{d\phi_c}{dx}\;.
\label{zerom}
\ee
From fig.(\ref{potential}-a) one can see that $\frac{d\phi_c}{dx}$
is zero only in the limit $x\rightarrow\pm\infty$, that is, the eigenfunction
$\psi_0$ has no nodes and then $\omega_0^2=0$ is the lowest eigenvalue. 
Then all the $\omega$'s are real and consequently the kink is stable
when quantum corrections are taking into account. On the other hand, from fig.
(\ref{potential}-b) one can see that  $\frac{d\phi_c}{dx}$ is zero for some 
finite $x=x_0$ in the turnig point $4$ (we can always choose this point  
by traslational invariance as corresponding to $x_0=0$). In this case the
eigenfuntion $\psi_0$ has a node, and then $\omega_0^2$ is
not the lowest eigenvalue. There is one negative eigenvalue $\omega^2<0$
and then one imaginary $\omega$. In this situation, the lump becomes
unstable by quantum corrections. In this case the eq.(\ref{massa}) 
has no direct physical interpretation, but the imaginary part signalize 
decay of a false vaccum \cite{coleman2}.

In both cases, to go further in the quantization program, we
have to solve a one dimensional Schrodinger equation but in general cases
this equation can not be analyticaly solved. 
 Instead to try to solve general Schrodinger equations, we
can adopt a different approach. We can start from a exactly solvable
Schrodinger equation to obtain the field theory model associated with it. 
The steps in this program are the following: first, from eq.(\ref{zerom}) 
solve it for $\phi_c$
\be
\phi_c(x)\propto\int^{x}\psi_0(y)dy\;.
\label{xsolve}
\ee
the second one is to invert for $x$ from eq.(\ref{xsolve}) obtaining $x=x(\phi_c)$. 
Thirth, we substitute eq.(\ref{zerom}) in eq.(\ref{bogo}) to obtain $U(\phi_c)$, that is
\be
U(\phi_c)=\frac{1}{2}\left(\frac{d\phi_c}{dx}\right)^2\propto\frac{1}{2}(\psi_0(x(\phi_c)))^2
\;.
\label{potencial}
\ee
Finally we can remove the subscript 'c', obtaining in this way the scalar field theoretic model. There are some points that we would like to stress.
From figures (\ref{potential}-a) and (\ref{potential}-b) we see that in principle
we only obtain in this way the part of $U(\phi)$ for $\phi$
that lies between the points $1$ and $2$ in the case of kinks
or in the case of lumps for $\phi$ that lies between the points $3$ and $4$.
Out of this intervals  in principle $U(\phi)$ obtained in this way can be any arbitrary 
function. From the above discussion there are  an infinite number of $U(\phi)$. Since
we are interested in field theory models that are smooth functions of $\phi$, the number
of possible $U(\phi)$ will reduce to one or zero. 

Before analyze particular Schrodinger potentials we would like to clarify
the reason  when and why we can remove the subscript 'c' in eq.(\ref{potencial}).
There are two different cases. The first one is the case where from eq.(\ref{potencial})
which is valid in principle only for those values of $\phi=\phi_c$, we use the same expresion 
for all values of $\phi$ that are out of $\phi_c$, and if the expression is still 
well defined for those values of $\phi$ then we have obtained a unique
and well behaved field theoretic model. The second situation is when eq.(\ref{potencial})
is not valid for those values of $\phi$ that lies out of $\phi_c$ (For example
(\ref{potencial}) would be imaginary or there will appear some singularities). Consequently
in this case there is no a well defined field theoretical model. 

As we have noted the eigenfuntion $\psi_0(x)$ as given by eq.(\ref{zerom}) is not
normalized. That is, we are obtaining the field
theoretical models modulo the coupling constants. With the functional form
of $U(\phi)$ in hands we can choose the coupling constants. 

In next sections we will construct field theory models starting from the following
integrable Schrodinger equations with three different potentials \cite{cooper}:\\
Rosen-Morse II hyperbolic
\be
A^2+B^2/A^2-A(A+1)/\cosh^2(x)+2B\tanh(x)\;,~~~~B<A^2\;,
\ee
the Morse potential:
\be
V(x)=A^2+B^2\exp(-2x)-2B(A+1/2)\exp(-x)
\ee
and finally the Scarf II hyperbolic potential:
\be
V(x)=A^2+(B^2-A^2-A)/\cosh^2(x)+B(2A+1)\tanh(x)/\cosh(x)\;.
\ee

\section{The Rosen-Morse II hyperbolic potential}

As we discused in the in the last section, the Rosen-Morse II hyperbolic potential
is given by
\be
V(x)=A^2+B^2/A^2-A(A+1)/\cosh^2(x)+2B\tanh(x)\;,~~~~B<A^2\;,
\label{rosen}
\ee
where $A$, $B$ are constants. For this potential the eigenfuntions are given by
\cite{cooper}
\be
\psi_n(x)=(1-y)^{\alpha/2}(1+y)^{\beta/2}~P_n^{(\alpha,\beta)}(y)\;,
\label{erosen}
\ee
with eigenvalues
\be
\omega_n^2=A^2-(A-n)^2+\frac{B^2}{A^2}-\frac{B^2}{(A-n)^2}\;, ~~~~n=0,1,2..
\ee
In eq.(\ref{erosen}) we have  $\alpha=A-n+B/(A-n),~\beta=A-n-B/(A-n)$, $y=\tanh(x)$
and $P_n^{(\alpha,\beta)}(y)$ are Jacoby polinomials \cite{abramovitz}. To reconstruct
field theory models that support kink-like solutions (stable models) from this potential
have to work with the ground state, the zero node
eigenfuntion $\psi_0(x)$. On the other hand, to reconstruct field theoretic models that
support lump-like solutions we have to consider the eigenfuntion with one node $\psi_1(x)$. 

\subsection{Stable Models}
To obtain the  kinks from the potential given by eq.(\ref{rosen})  
we have to integrate the ground state eigenfuntion $\psi_0(x)$ that 
can be obtained from eq.(\ref{erosen}),
\be
\psi_0(x)=(1-y)^{\alpha/2}(1+y)^{\beta/2}\;.
\ee
Using eq.(\ref{zerom}) we obtain the kink
\be
\phi_c(x)=\int^{\tanh(x)}\frac{(1-y)^{\alpha/2}(1+y)^{\beta/2}}{1-y^2}dy\;,
\label{rkink}
\ee
and from eq.(\ref{potencial}) we obtain for $U(\phi_c)$
\be
U(\phi_c)=(1-y)^{\alpha}(1+y)^{\beta}\;.
\label{rpotencial}
\ee
Note that in general the integral given by eq.(\ref{rkink}) can not be performed analyticaly. Consequently, we will be restrict the following cases:\\

\underline{{\bf 3.1.a~~$\alpha=\beta=A$}}~: In this case $B=0$, 
and eq.(\ref{rkink}) becomes
\be
\phi_c(x)=\int^{\tanh(x)}\frac{(1-y^2)^{A/2}}{1-y^2}dy\;.
\label{rkink2}
\ee
The above integral can ben performed in terms of elementary functions only
when $A$ is an integer. In eq.(\ref{rpotencial}) we have in principle\
that $-1\leq y\leq 1$, (this follows from the definition $y=\tanh(x)$). 
In this interval we only obtain $U(\phi)$ for those values that take $\phi_c$. 
As was mentioned in last section, to obtain $U(\phi)$ for all $\phi$
we have to extend the eq.(\ref{rpotencial}) for all values of $\phi$
and see if it is still well behaved. For integer $A$ we see that eq.(\ref{rpotencial}) is a well behaved funtion for all values of $y$.
Then extending eq.(\ref{rpotencial}) for all $\phi$ is equivalent to 
extend it for all $y$. Then if we invert eq.(\ref{rkink2}) for $y=\tanh(x)$ as
function of $\phi_c$, extend such result for all values of $\phi$ 
(and then for all values of $y$) in a well behaved 
way and replacing it in eq.(\ref{rpotencial}) we will obtain a well behaved field
theory model. But for $A>2$ from eq.(\ref{rkink}) we can solve $d\phi/dy=(1-y^2)^{(A-2)/2}=0$ and we find that the points $y=\pm 1$ 
signalizes maxima or minima for 
$\phi$ as function of $y$,
that is $\phi$ as function of $y$ is not singled valued, then it will
be not possible to invert $y$ as function of $\phi$ in unique way. One can
try to circumvent this difficulty supressing those values of $y$ in 
which $\phi$ is not single valued. Nevertheless, we will be able
only to obtain $U(\phi)$ by parts, generating in this way discontinuities, for example in the derivatives of $U(\phi)$. 
The same type of singularities has been noted in \cite{lee}. For $A<2$ {\it i.e}, $A=1,2$ this problem does not occur. One can perform the integral given by eq.(\ref{rkink2}) and replace in eq.(\ref{rpotencial}) to obtain the well know sine-Gordon and $\phi^4$ models respectively.

We would like to stress the following point. We are considering the 
case in which $V(x)=-A(A+1)/\cosh^2(x)$ for $A$ integer. Such potentials 
have both, discrete and continuous modes with the advantageus 
property  of being reflectionless. The first quantum corrections to the mass
of the kinks are given by eq.(\ref{massa}). To sum the continuous modes,
we have to know the density of states. For the case of reflectionless
potentials this can be given in terms of the phase shifts of the 
one dimensional scattering problem.  In general this sum is logaritmically
divergent and we need to renormalize the theory. In two-dimensional scalar
field theories such divergences can be eliminated using a normal ordering
prescription. These properties was used by Cahill {\it et al} \cite{cahill} to
find a finite result for
the quantum corrections to the mass of the static solitons. Moreover
they also obtained the quantum mass corrections only in terms of the discrete eigenvalues of the associated Schrodinger 
equation. Using the former result Boya {\it et al} \cite{prd41-1990} 
obtained a closed expression for the first quantum corrections for 
the mass of the static
kinks given by eq.(\ref{rkink2}) for any integer $A>0$ without explicit
knowledgement of the
field theory model that support such kinks. But here a question arises. Are
these expressions valid even we have showed that there is not possible to construct a well
behaved field theory model for $A>2$? It is possible to show that these
expressions are valid since the kinks only see
the parts of the field theory models that lies between the degenerate vacua.
In such domain, that is, between the extremum values of $\phi_c$ it is allways
possible to invert $y$ as function of $\phi$, and then it will be possible
to reconstruct the field theory model for those values of the field that lies
between the degenerate vacua. Out of this interval, the theory can be anything. 
In other words the complete field theory model is ambiguous, but in general not well behaved. We can say that the masses of the kinks (for $A>2$) that was computed in \cite{prd41-1990} are the masses of an infinite class of not well behaved field theory models.\\

\underline{{\bf 3.1.b~~$\alpha\neq\beta$}}~: In this case $B\neq 0$. We can
rewrite eq.(\ref{rkink}) as
\be
\phi_c(x)=\int^{\tanh(x)}(1-y)^{(\alpha-2)/2}(1+y)^{(\beta-2)/2}dy\;.
\label{rkink4}
\ee
Choosing $(\alpha-2)/2=m$ and $(\beta-2)/2=n$ we have
\be
\phi_c(x)=\int^{\tanh(x)}(1-y)^n(1+y)^m dy\;.
\label{rkink5}
\ee
Let us consider the case in which $n=0$. Then, from eq.(\ref{rkink5}) we 
obtain
\be
\phi_c(x)=(1+\tanh(x))^{m+1}
\ee
from which we can solve for $y=\tanh(x)$ and then substituting in eq.(\ref{rpotencial}) we obtain
\be
U(\phi)=\phi^2\left( 2 -\phi^{1/(m+1)}\right)^2\;.
\ee
We see that for the values of $m$ such that $1/(m+1)$ is fractionary
we will have in some cases (for example when  $1/(m+1)=1/2$) an imaginary value for $U(\phi)$. 
For such values of $m$ we can redefine $\phi^{1/(m+1)}$ as $(\phi^2)^{1/2(m+1)}$ to make 
$U(\phi)$ a real valued funtion, but in these cases we obtain a discontinuity
in the derivative for $U(\phi)$, making the theory  not well behaved. If we
take $m$ such that $1/(m+1)=2l$ with $l$ integer we will obtain
\be
U(\phi)=\phi^2\left( 2 -\phi^{2l}\right)^2\;,
\label{mm}
\ee
that is, we obtain a well behaved polinomial like field theory models with
three degenerate vacua. The case $l=1$ is the $\phi^6$ model with three
degenerate vacuum that was considered in ref. \cite{lohe}. In this paper Lohe obtained
the expression of the renormalized mass of the soliton. It is interesting to point out that Flores {\it et al} \cite{flores}
studying the vacuum decay rate in the the massive $\phi^6_{3D}$ model in the thin wall
approximation obtained the same kink solution associated with the model given by the $l=1$ 
case. Note that in this case  $V(x)$ given by eq.(\ref{rosen})
is not reflectionless making the computations of the quantum mass corrections
very hard. The eq.(\ref{mm}) also was analyzed in \cite{cas2}. If we put $1/(m+1)=2l+1$ 
with $l$ integer we will obtain 
\be
U(\phi)=\phi^2\left( 2 - \phi^{2l+1}\right)^2\;,
\label{ll}
\ee
that is, well behaved field theory models with two degenerate vacua. The case
$l=1$ is the case of the $\phi^8$ theory with two degenerate vacua. The
above field theory models was also obtained in  ref. \cite{cas2}. If we
consider the case in which $m=0$ we obtain
the same configuratios that in the $n=0$ case. If we consider the cases
in which both $n$ and $m$ are integers, we can still integrate the eq.
(\ref{rkink5}) but in in this case it is not possible to obtain  a well 
behaved field theory model for the same reason stressed in the $B=0$ case.
From the above discussion we conclude that these are the only cases in which
we can reconstruct well behaved stable field theoretic models. In the next sub-section we will analyze the unstable models (lumps).

\subsection{Unstable models}
In this case the lumps are obtained integrating the $n=1$ case of eq.(\ref{erosen})
\be
\psi_1(x)=(1-y)^{\alpha/2}(1+y)^{\beta/2}\left((\alpha+\beta+2)y+\alpha-\beta\right)\;,
\label{erosenf}
\ee
that is we have for the bounces
\be
\phi_c(x)=\int^{\tanh(x)}\frac{(1-y)^{\alpha/2}(1+y)^{\beta/2}
\left((\alpha+\beta+2)y+\alpha-\beta\right)}{1-y^2}dy\;.
\label{rosboun}
\ee
The field theoretic model are given by
\be
U(\phi)=(1-y)^{\alpha}(1+y)^{\beta}\left((\alpha+\beta+2)y+\alpha-\beta
\right)^2\;.
\label{inespot}
\ee
It is possible to analytically solve the integral given by eq.(\ref{rosboun})
in the following cases\\

\underline{{\bf 3.2.a~~$\alpha=\beta=A$}}~: In this case $B=0$. Consequently
we have for $\psi_1(x)$
\be
\psi_1(x)=2A\frac{\sinh(x)}{\cosh^{A}(x)}\;,
\label{Bocase}
\ee
and the integral in eq.(\ref{rosboun}) can be easily performed. We obtain 
\be
\phi_c(x)=\frac{1}{\cosh^{A-1}(x)}\;.
\label{urg}
\ee
Substituting the eq.(\ref{urg}) in eq.(\ref{inespot}) we obtain the field theory models
\be
U(\phi)=(A-1)\phi^2(1-\phi^{2/(A-1)})\;.
\label{mrst}
\ee
For $A=2$ we have the unstable $\phi^4$ theory. This model was used
by Langer \cite{langer} as a field theoretic model for the study of the kinematics
of first order phase transitions. Further, Coleman and Callan \cite{coleman2} extended
the Langer's work to the related issue of the false vacuum decay in field theory \cite{coleman2}.
For $A=3$ we obtain the unstable $\phi^3$ model. This last model
was used as a laboratory for computing the decay rate of a system trapped in a false vacuum \cite{IJMPA-5-1319}. Also, recently
the $\phi^3$ model was used as an exactly solvable toy model for tachyon
condesation in string field theory \cite{zwiebach}. 
For some values of $A$ such that $2/(A-1)$ is fractionary the eq.(\ref{mrst}) not describe
a well behaved field theoretic model.
Also we have to stress that we must assume $A>0$ to garantee
the normalizability of $\psi_1(x)$. In ref. \cite{cas2} the authors considered the case $A=1$ and obtained the Liouville field theory model \cite{liou}.
But in such case is easy to see that the classical solutions that meet or leave the unique 
asymptotic vacuum have an infinite energy, that is, such solutions are not lumps. Finally
that we would stress that the density potential given by eq.(\ref{mrst}) was also obtained
in references \cite{cas2} and \cite{zwiebach}.\\
\underline{{\bf 3.2.b~~$\alpha\neq\beta$}}~: In this case $B\neq 0$ and  
the integral given by eq.(\ref{rosboun}) splited in two parts,
\be
\phi_c(x)=(\beta+\alpha+2)\int^{\tanh(x)}\frac{(1-y)^{\alpha/2}(1+y)^{\beta/2}y}{1-y^2}dy+(\alpha-\beta)\int^{\tanh(x)}\frac{(1-y)^{\alpha/2}(1+y)^{\beta/2}}{1-y^2}dy\;.
\label{fazer}
\ee
The above integral can be performed for some particular values of $\alpha$, $\beta$
but it is not possible to obtain a well behaved field theoretic model.

\subsection{The $A\rightarrow \infty$ Limit}
In ref. \cite{minahan} the limit $A\rightarrow\infty$ of eq.(\ref{mrst}) was studied.
It was obtained the unstable field theoretic model
\be
U(\phi)=-\phi^2\ln\phi^2
\ee
with lumplike solution given by
\be
\phi_c(x)=x\exp(-x^2/2)\;.
\ee
It is easy to see that in this case the Rosen-Morse potential given by eq.
(\ref{rosen}) reduces to one of the harmonic oscillator type. To perform the 
limit first one rescale the $x$ 
coordinate that appear in eq.(\ref{rosen}) as $x/\sqrt{2(A-1)}$. The authors
considered first the case in which $A$ is integer and then performed the
$A\rightarrow\infty$ limit. Recently it was considered \cite{hep-th/0104229}
the case in which $A$ is arbitrary, and was showed that the
$A$ infinite limit can be taken continuosly. But as we showed in last
subsection the eq.(\ref{mrst}) does not define a well behaved theory for those 
values of $A$ as for example $2/(A-1)=1/2$. In other words one can not
take the limit $A\rightarrow\infty$ continuosly.

Using the harmonic oscillator potential the authors of ref. \cite{minahan2} 
also considered the reconstruction of the stable
field theoretic model. In this case it is not possible to obtain a closed 
expression of $U(\phi)$ since to obtain the kink we have to
integrate $\exp(-x^2/2)$. Unfortunately one can not express the result in terms of elementary
funtions. Indeed it is given by a error funtion, that can not be inverted in terms of $\phi_c$.
In order to circumvent this problem the authors of ref. \cite{minahan2} redefined
the field to write an explicit form for the lagrangean of
the field theoretic model. At this point an interesting question arises. We  showed in the last section
that it is not possible to reconstruct stable field theory models, with
exception of the $A=1$ and $A=2$ cases.  Is the model considered in ref. \cite{minahan2}
a well behaved field theoretic model? Note that we  analyzed  only the cases
of finite values of $A$. For $A\rightarrow\infty$ our analysis is incomplete.
The answer to the above question is afirmative. The argument for it is
the following: in the models that we studied the singularities always
appear in the perturbative vacua, $\phi=\phi_0$. We can solve for $U(\phi)$ for any value of $A$ around one of the trivial vacua as has been done in ref. \cite{lee}. The result is
\be
U(\phi)\approx (\phi-\phi_0)^2+{\mathcal O}\left((\phi-\phi_0)^{2+2/A}\right)\;.
\ee
In the limit in which $A\rightarrow\infty$ we see that the singularity
in $\phi=\phi_0$ disappear, showing that in this limit 
a well behaved field theory model is obtained. It is important to point out the following. Redefining the field, the authors of ref. \cite{minahan2} obtained 
the following density lagrangean
\be
{\mathcal L}\sim \left(\frac{1}{2}e^{-2T^2}\partial_\mu T\partial^\mu T-\frac{1}{8}
e^{-2T^2}\right)\;,
\label{minahan}
\ee
where $T$ is the redefined field given by
\be
T={\mathrm erf}^{-1}(\phi)\;,
\ee
and the kink is given by
\be
\phi_c(x)={\mathrm erf}(x/2).
\ee
Note that the lagrangean density writen in terms of the new field
$T$ does not cover all the values of the field $\phi$, since the error
funtion is a funtion with finite range. The inverse 
funtions will be defined only for a finite domain, {\it i.e} for those
values of $\phi$ that lies between the trivial vacua. 
Consequently, althought
the $A$ infinite limit defines a well behaved field theoretic model
the density lagrangean given by eq.(\ref{minahan}) does not describe the complete theory. For example it is not valid to perform the perturbative analysis of the model, althought for the solitonic sector it is still useful.

\section{The Morse potential: the $\frac{m^2}{8}\phi^2\ln^2\left(\frac{\alpha^2\phi^2}
{9m^4}\right)$ model.}
The Morse potential is given by 
\be
V(x)=A^2+B^2\exp(-2x)-2B(A+1/2)\exp(-x)\;.
\label{morse}
\ee
In this case the eigenfuntions and eigenvalues of eq.(\ref{morse}) are
given by
\be
\psi_n(x)=y^{A-n}e^{-y/2} L_n^{2A-2n}(y),~~~y=2Be^{-x}\;,
\label{eigenm}
\ee
and
\be
\omega_n^2=A^2-(A-n)^2,~~~~n=0,1,2..\;.
\label{autom}
\ee
From eq.(\ref{eigenm}) we obtain the ground state eigenfuntion $\psi_0$
\be
\psi_0(x)=y^Ae^{-y/2}\;,
\label{eigenm2}
\ee
from which we can obtain the kink
\be
\phi_c(x)=\int^{2Be^{-x}}y^{A-1}e^{-y/2}dy\;.
\label{kinkm1}
\ee
The stable field theory model is given by
\be
U(\phi)=y^{2A}e^{-y}\;.
\label{potenmor}
\ee
The integral given by eq.(\ref{kinkm1}) can be performed only when $A=1,2,..$ For $A=1$
we obtain
\be
\phi_c(x)=\exp(-Be^{-x})\;,
\ee
from which, solving for $x$ and replacing in eq.(\ref{potenmor})
we obtain
\be
U(\phi)=\phi^2(\ln(\phi^2))^2\;.
\label{new1}
\ee
Although the integral given by eq.(\ref{kinkm1}) can be done
for $A=2,3..$
it is not possible to invert it to obtain a well behaved field theory model. 
In this case it is not possible to reconstruct well behaved unstable
field theory models. 

\begin{figure}
\centerline {\epsfxsize=3in\epsfysize=1.5in\epsffile{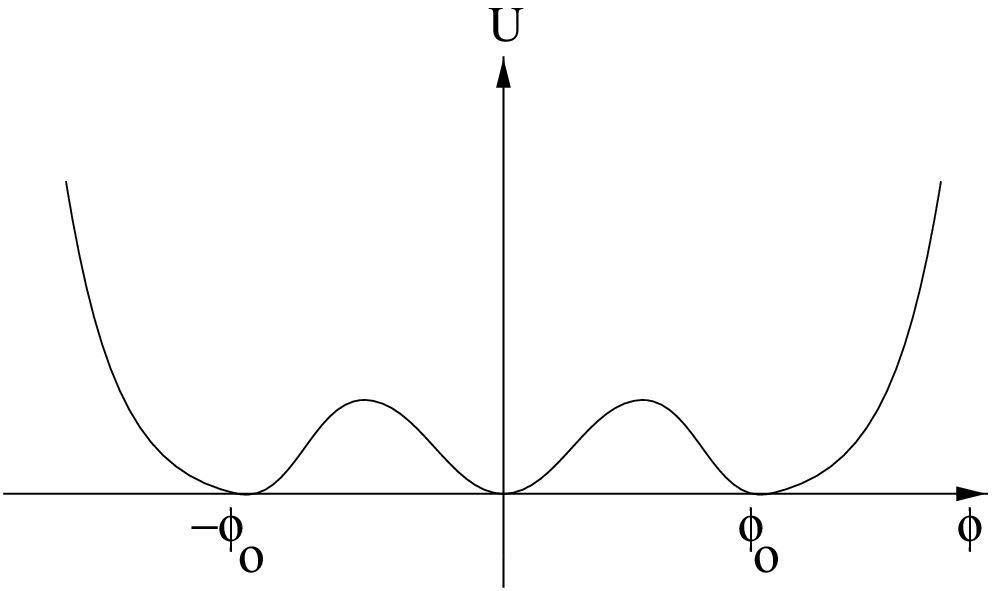} }
\caption{The density potential $U(\phi)$ given by eq.(\ref{potm}).}
\begin{picture}(10,10)
\end{picture}
\label{morsef}
\end{figure}

The density potential given by eq.(\ref{new1}) can be redefined in such a way
that will appear coupling constants in the model. Since we constructed this
field theory model starting from a Schrodinger equation with one free parameter ($A=1$ and $B$ arbitrary)
and since we can  rescale field and coordinates in the lagrangean 
(thus eliminating two coupling constants) we conclude that the
density potential
given by eq.(\ref{new1}) can be redefined with no more than three coupling constants.
We redefine eq.(\ref{new1}) with two coupling constants as
\be
U(\phi)=\frac{m^2}{8}\phi^2\ln^2\left(\frac{\alpha^2\phi^2}
{9m^4}\right)\;,
\label{potm}
\ee
where we have choose the numerical factors adequately. The model given by eq.(\ref{potm}) have two degenerate vacua
as can be see in fig. (\ref{morsef}) at the points given by $\pm \phi_0$,
where
\be
\phi_0=\frac{3m^2}{\alpha}\;.
\ee
The kinks and antikinks can be obtained easily, they are given by
\be
\phi_c(x)=\pm\frac{3m^2}{\alpha}\exp(-e^{\pm m(x-x_0)})\;.
\ee
We have two pair of kink antikink solutions that can link $\phi=-\phi_0$
and $\phi=0$ or $\phi=0$ and $\phi=\phi_0$. The masses of these kinks (antikinks)
solutions are the same and are given (classically) by   
\bqn
H[\phi_c]&=&\int_{-\infty}^{\infty}dx\left (\frac{1}{2}\left(\frac{d\phi_c}{dx}\right)^2+
\frac{m^2}{8}\phi_c^2\left[\ln\left(\frac{\alpha^2\phi_c^2}
{9m^4}\right)\right]^2\right)\nonumber\\
&=&\frac{9m^6}{2\alpha^2}\int_{-\infty}^{\infty}dxe^{\pm mx}\exp(-2e^{\pm mx})
\nonumber\\
&=&\frac{9m^5}{4\alpha^2}\;.
\eqn

Let us briefly develop how to find the quantum corrections
for the soliton mass. As we discussed this quantity is given by eq.(\ref{massa}). To define the soliton mass  we have to substract from eq.(\ref{massa})
the energy fluctuations around the asymptotic limits of the solitonic solution,
that is, around the perturbative vacua $\phi=\pm\phi_0$. This subtraction
only modify the quantum corrections since in the perturbative vacuum 
$H[\pm\phi_0]$ vanishes. With this modification we have for the mass of
the kinks
\be
M=H[\phi_c]+\sum_n\omega_n-\sum_q\omega_0(q)\;.
\label{qmass}
\ee
In eq.(\ref{qmass}) the $\omega_n$'s satisfy eq.(\ref{estab}), that in the present case is
\be
\left[-\frac{d^2}{dx^2}+m^2(e^{\pm 2mx}-3e^{\pm mx}+1)\right]\psi_n(x)=\omega_n^2\psi_n(x)\;,
\label{re}
\ee
where $\pm$ signs correspond respectively to the kink and antikink
configurations. On the other hand $\omega_0(q)$ satisfy
\be
\left[-\frac{d^2}{dx^2}+V(x)\right]\psi_n(x)=\omega_n^2\psi_n(x)
\label{ra}
\ee
with $V(x)$ given by the asymtotic behavior of the potential in eq.
(\ref{re}) at $x\rightarrow\pm\infty$. If we kept the $(-)$ sign in
eq.(\ref{re}) then $V(x)$ is given by
\begin{displaymath}
V(x)=\left\{\begin{array}{ll}
\infty & x\leq 0\\
m^2 &x>0\end{array}\right.
\end{displaymath}
If we kept the $(+)$ sign the form of $V(x)$ is reverted. But in any
case the quantum mass correction for the kinks or antikinks will be
the same. If in eq.(\ref{re}) we make $z=mx$ we obtain
\be
\left[-\frac{d^2}{dz^2}+e^{\pm 2z}-3e^{\pm z}+1\right]\psi_n(z)=\delta_n\psi_n(z)\;,
\label{re1}
\ee
with $\delta_n=\omega_n/m^2$. Note that eq.(\ref{re1}) is the $A=B=1$ case
of the Morse potential as espected by construction.

The quantum mass correction for the soliton as given by eq.(\ref{qmass})
is in general divergent. In order to renormalize it we have to make
a redefinition af the parameters of the theory. To carrie such task first
we expand $U(\phi)$ around one of the perturbative vacua,
\be
U(\varphi)=\frac{1}{2}m^2\varphi^2\pm\frac{\alpha}{6}\varphi^3-
\frac{1}{216}\varphi^4+\frac{1}{14580}\frac{\alpha^4}{m^6}\varphi^6+...
\label{taylor1}
\ee
where we have shifted the field as $\phi=\varphi\pm\phi_0$. Since we are
in the semi-classical aproximation we will renormalize the mass $m$ and
the coupling constant to one loop order. Details of the renormalization
procedure and the calculation of the quantum correction to the soliton mass
in this model will be present elsewhere \cite{novo}.

\section{The Scarf II Hyperbolic potential: the $\frac{m^2}{8}\phi^2\cos^2 \left[ \ln\left(\frac{\alpha^2\phi^2}{9m^4}\right)\right]$ model.}
The Scarf II hyperbolic potential is given by
\be
V(x)=A^2+(B^2-A^2-A)/\cosh^2(x)+B(2A+1)\tanh(x)/\cosh(x)\;.
\label{scarf}
\ee
In this case the eigenfuntions and eigenvalues of this potential are
given by
\be
\psi_n(x)=(i)^n(1+y^2)^{-A/2}e^{-B\tan^{-1}(y)}
P_n^{(iB-A-1/2,-iB-A-1/2)}(y),~~y=\sinh(x)\;,
\label{eigenscar}
\ee
and
\be
\omega_n^2=A^2-(A-n)^2~,~~~n=0,1,2...
\label{autscarf}
\ee
The kink like solutions are obtained from the zero mode given by
\be
\psi_0(x)=(1+y^2)^{-A/2}e^{-B\tan^{-1}(y)}\;,
\label{lowsc}
\ee
and the field theory model is given by
\be
U(\phi)=(1+y^2)^{-A}e^{-2B\tan^{-1}(y)}\;.
\label{potenscar}
\ee
Integrating the eq.(\ref{lowsc}) we obtain for $\phi_c$
\be
\phi_c(x)=\int^{\sinh(x)}(1+y^2)^{-(A+1)/2}e^{-B\tan^{-1}(y)}dy\;.
\label{kinkscar}
\ee
The above integral can be performed analyticaly only when $A=1$.
We have 
\be
\phi_c(x)=e^{-B\tan^{-1}\sinh(x)}\;,
\label{kinks2}
\ee
then solving for $y=\sinh(x)$ and replacing in eq.(\ref{potenscar})
we obtain 
\be
U(\phi)=\phi^2\cos^2(\frac{1}{2B}\ln\phi^2)\;.
\label{new2}
\ee
As in the case of the Morse potential in this case we not have
unstable field theory models.

\begin{figure}
\centerline {\epsfxsize=3.2in\epsfysize=1.8in\epsffile{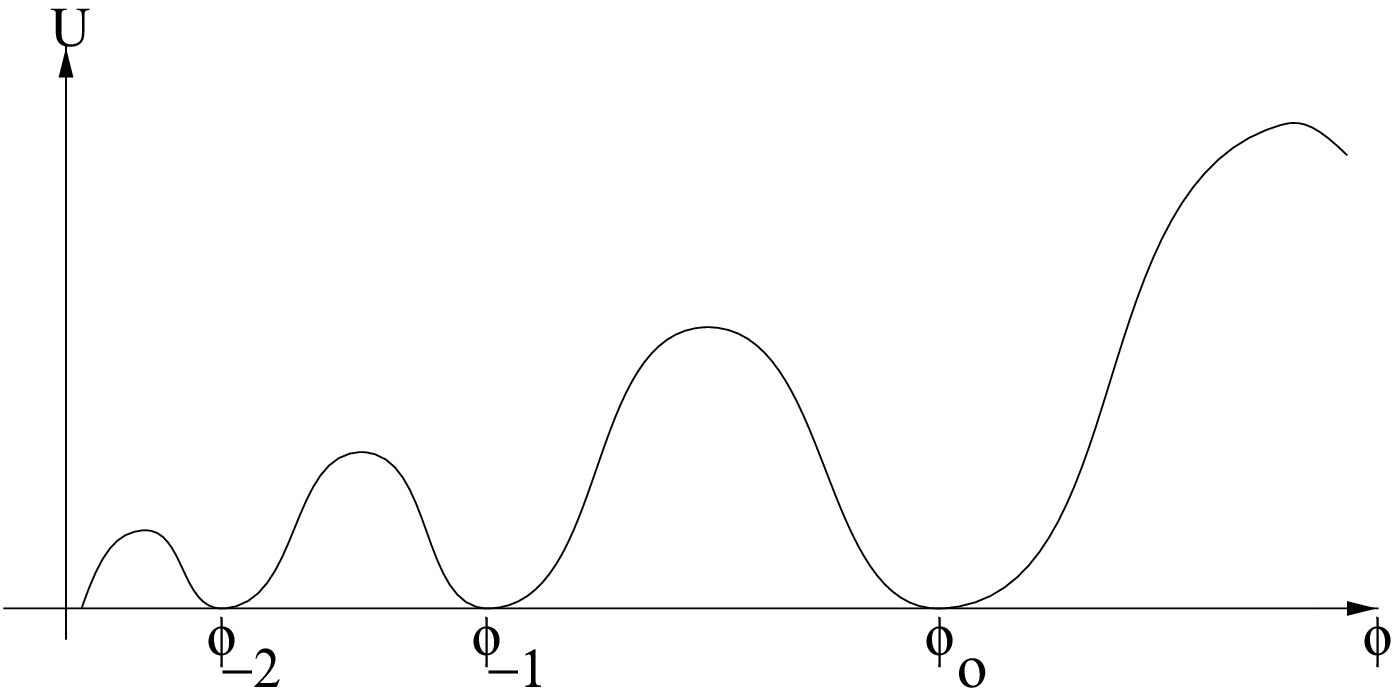} }
\caption{The density potential $U(\phi)$ given by eq.(\ref{new3}).}
\begin{picture}(10,10)
\end{picture}
\label{scarff}
\end{figure}

We can redefine the potential given by eq.(\ref{new2}) with adequate
coupling constants (we consider the most simple $B=1/2$ case)
\be
U(\phi)=\frac{m^2}{8}\phi^2 \cos^2 \left[ \ln\left(\frac{\alpha^2\phi^2}{9m^4}\right)\right]\;.
\label{new3}
\ee
In fig. (\ref{scarff}) we have plotted this density potential for $\phi>0$. It has infinitely degenerate trivial vacua at the points $\phi=\pm\phi_n$
with $\phi_n$ given by
\be
\phi_n=\frac{3m^2}{\alpha}\exp\left(\frac{2n+1}{4}\pi\right)\;, ~~n=0,\pm 1,\pm 2,..
\label{zeros2}
\ee

The kinks and antikinks are obtained using eq.(\ref{classic})
\be
\phi_c(x)=\pm\frac{3m^2}{\alpha}\exp\left(\frac{n\pi}{2}
 \pm\frac{1}{2}\tan^{-1}(\sinh(mx))\right)\;,~~n=0,
\pm 1,\pm 2,..
\ee
where the solutions with $(\pm)$ signs in the exponents correspond to the kinks anti kinks solutions respectively for each value of $n$ and for each sign
that appear in front. We have an infinite number os kinks and anti-kinks
that links the infinite number of trivial vacua. This makes remember in some
sense the sine-Gordon model. But contrary to sine-Gordon model where
all the solitonic sectors describe the same physics in our present case it is not the case. For example
if we compute the the classical masses for the kinks (or anti-kinks) we obtain
\bqn
H[\phi_c]&=&\int_{-\infty}^{\infty}\left(\frac{1}{2}\left(\frac{d\phi_c}{dx}\right)^2+\frac{m^2}{8}\cos^2\left[\ln\left(\frac{\alpha^2\phi^2}{9m^4}\right)\right)\right]+\nonumber\\
&=&\frac{9m^5e^{2n\pi}}{4\alpha^2}\int_{-1}^{1}ds e^{\sin^{-1}(s)}\nonumber\\
&=&5.65\frac{m^5e^{2n\pi}}{\alpha^2}\;,
\eqn
where we see clearly that the masses are different. The quantum mass corrections
are given by the last two terms of eq.(\ref{qmass}) where now
the $\omega_n$'s satisfy
\be
\left(-\frac{d^2}{dx^2}+m^2-(7/4)m^2/\cosh^2(mx)+(3/2)\tanh(x)/\cosh(x)\right)
\psi_n(x)=\omega_n^2\psi_n(x)
\label{scarf5}
\ee
and the $\omega_0(q)$'s satisfy the eq.(\ref{scarf5}) at asymtotic values,
that is, an equation with constant potential equal to $m^2$. As usual
the quantum mass correction for the soliton mass is divergent and in order to
renormalize it we have to expand eq.(\ref{new3}) around one of
the perturbative vacua 
\be
U(\varphi)=\frac{m^2}{2}\varphi^2\pm\frac{\alpha}{6}e^{-(2n+1)\pi/4}\varphi^3
-\frac{17}{9(4!)}\frac{\alpha^2}{m^2}e^{-(2n+1)\pi/2}\varphi^4
+\frac{340}{6!81}\frac{\alpha^4}{m^6}e^{-5(2n+1)\pi/4}\varphi^6+...
\label{tayscar}
\;.
\ee
Where we have redefined $\phi=(\varphi\pm\phi_n)$ with $\phi_n$ given by
eq.(\ref{zeros2}). We can see in eq.(\ref{tayscar}) that when $n\rightarrow\infty$ the theory becomes free in the perturbative sector,
although this fact does not happens in the solitonic sector \cite{novo}.

\section{Conclusions}
Using the zero mode solution of the  Schrodinger type 
equation for kinks (bounces) we obtain the kinks (bounces) as well the
bidimensional scalar field theory models that support such kinks (bounces).
Because we start with solvable Schrodinger equations it is automatically
guaranted the computation of the first quantum corrections around the
kinks (bounces).  We obtained two 
new models starting from the Morse and Scarf II hyperbolic potentials,
{\it i.e}, the $U(\phi)=\phi^2\ln^2(\phi^2)$ and $U(\phi)=\phi^2\cos^2(\ln(\phi^2))$. The quantum corrections to the
solitonic sectors will be presented elsewhere. 

\vspace{1.5cm}
{\bf Acknoledments}:\\

This paper was supported by Conselho Nacional de Desenvolvimento 
Cientifico e Tecnologico (CNPq) of Brazil.

\newpage

\end{document}